\documentclass[12pt]{article}

\def\tr{\mathop{\rm tr}}
\def\im{\mathop{\rm Im}}
\def\re{\mathop{\rm Re}}

\begin{document}
\title{Spectral Statistics of the $k$-Body Random-Interaction
Model}

\author{
Mark Srednicki \\
Department of Physics \\ 
University of California \\
Santa Barbara, CA 93106 USA}

\date{July 8, 2002}

\maketitle

\begin{abstract}
We reconsider the question of the spectral statistics
of the $k$-body random-interaction model, investigated
recently by Benet, Rupp, and Weidenm\"uller,
who concluded that the spectral statistics are Poissonian.
The binary-correlation method that these authors used
involves formal manipulations of divergent series.
We argue that Borel summation does not suffice to 
define these divergent series without further
(arbitrary) regularization, and that this constitutes
a significant gap in the demonstration of Poissonian
statistics.  Our conclusion is that the spectral
statistics of the $k$-body random-interaction model
remains an open question.

\end{abstract}

\newpage

\section{Introduction}
\label{int}

In recent work, Benet, Rupp, and Weidenm\"uller 
(hereafter BRW) \cite{brw1,brw2}
have considered the spectral statistics of $k$-body random
interaction models \cite{fre71,boh71,mf}.
These are models of fermions (usually;
boson versions can also be studied \cite{abrw})
in which there are $\ell$ single-particle states occupied by
a total of $m$ particles.
The models are specified by hamiltonians of the form
\begin{equation}
H_k \ = \sum_{{1 \leq j_1 < j_2 < \ldots < j_k \leq \ell} \atop 
              {1 \leq i_1 < i_2 < \ldots < i_k \leq \ell}       } 
        V_{j_1 \ldots j_k , i_1 \ldots i_k}
  a_{j_1}^{\dagger} \ldots a_{j_k}^{\dagger} a_{i_k} \ldots a_{i_1} \ ,
\label{hk}
\end{equation}
where the $a^\dagger$'s and $a$'s are creation and annihilation
operators, and the $V$'s are (for the case of the unitary ensemble)
complex numbers that obey
\begin{equation}
\label{v}
V_{j_1 \ldots j_k, i_1 \ldots i_k} = 
V^*_{i_1 \ldots i_k, j_1 \ldots j_k} \ .
\end{equation}
Each independent component is a Gaussian random variable with
mean zero and variance $v_{0k}^2$,
\begin{equation}
\overline{V_{j_1 \ldots j_k, i_1 \ldots i_k}
    V_{j'_1 \ldots j'_k, i'_1 \ldots i'_k}} = 
    v_{0k}^2 \, \delta_{j_1 i'_1} \ldots 
    \delta_{j_k i'_k} \delta_{i_1 j'_1} \ldots
    \delta_{i_k j'_k} \ ;
\label{v0}
\end{equation}
the overbar denotes averaging over the ensemble,
and $v_{0k}$ is a normalization constant that sets the energy scale.
In this paper, for pedagogical simplicity, we will restrict 
our attention to the unitary ensemble.

These models can be viewed as caricatures of complex systems of
interacting particles, such as nuclei, multi-electron atoms, 
or quantum dots.  The most physically interesting case is $k=2$:
$H_2$ can be thought of as representing two-body interactions among
a set of $m$ particles occupying $\ell$ degenerate single-particle
states.

In their work, BRW analyzed the spectral statistics of these models,
and found that they are Poissonian for $\ell \gg m\gg k$.  
This is something of a surprise; numerical simulations of the 
$k=2$ model have generally found Wigner-Dyson spectral statistics
(see, e.g., \cite{brody,guhr}).
Even more surprising, the proof of
this offered by BRW extends trivially to the case of nondegenerate
single-particle levels specified by $H=H_1+H_2$, for {\it arbitrary\/}
values of $v_{02}/v_{01}$ \cite{brw3}.  

This is counter-intuitive on at least two levels.  
First, mathematically, one would naturally expect 
that determining the spectral statistics of $H_1+H_2$ would
be a much harder problem than it is for either $H_1$ or $H_2$ alone.
Second, physically, we expect this model to capture the essential physics
of a chaotic many-body system (with two-body interactions), just
as a single random matrix is often a good model of a
chaotic few-body system.  According to the Bohigas--Giannoni--Schmit
conjecture \cite{bgs} (see also \cite{cas}), a classically chaotic
system should exhibit Wigner-Dyson statistics at sufficiently high
energy (irrespective of the number of degrees of freedom).  
If the BRW result is correct, we must then conclude that
either the BGS conjecture is false for many-body systems, 
or that the two-body random-interaction model does not correctly capture
this aspect of the physics of these systems.

In this paper we will show that there is in fact a gap in the BRW proof, and
that this gap is not easily bridged.  Thus, we will argue, 
Wigner-Dyson statistics for these models is still an open possibility.

In section~2 we review and expand the analysis of BRW \cite{brw1,brw2},
and explain how their proof of Poissonian spectral statistics might fail
to hold.  In section~3 we specialize to the case $k=1$ (where the detailed
analysis simplifies) and show that, within the context of the 
binary correlation method used by BRW, the proof does break down in 
the manner suggested in section~2.    
However, the binary correlation method is not exact, and the appropriate 
conclusion is that one must go beyond this
approximation to obtain a reliable evaluation of
the spectral statistics of these models.
We elaborate on this further in Section~4.

\section{The binary correlation method}
\label{bnc}

We begin with the resolvent 
\begin{equation}
G(z) = \tr {1\over z-H} \ ,
\label{g}
\end{equation}
where $\tr$ denotes the normalized trace ($\tr 1=1$)
over states of $m$ particles.
BRW compute the connected correlation function
\begin{equation}
R(z_1,z_2) = \overline{G(z_1)G(z_2)} - 
             \overline{G(z_1)}\,\,\overline{G(z_2)}\ .
\label{r}
\end{equation}
Then, since the density of states $\rho(E)$ is given by
\begin{equation}
\rho(E) = {1\over2\pi i}\Bigl[G(E-i\varepsilon)-G(E+i\varepsilon)\Bigr]\ ,
\label{rho}
\end{equation}
where $E$ is real and $\varepsilon$ is a positive infinitesimal,  
we have
\begin{eqnarray}
\overline{\rho(E_1)\rho(E_2)} - \overline{\rho(E_1)}\,\,\overline{\rho(E_2)} 
&=& {1\over(2\pi)^2}\Bigl[R(z_1^+,z_2^-)+R(z_1^-,z_2^+) 
\nonumber \\
         && \qquad {}-R(z_1^+,z_2^+)-R(z_1^-,z_2^-)\Bigr] \ ,
\label{rhorho}
\end{eqnarray}
where $z_i^\pm = E_i\pm i\varepsilon$.
Thus, we can extract the connected density-density correlation 
(which provides information on the spectral statistics) from $R(z_1,z_2)$,
{\it provided\/} we can evaluate $R(z_1,z_2)$ with both $z_1$
and $z_2$ on either side of the real axis.

To compute $R(z_1,z_2)$, BRW expand in powers of $H$,
\begin{equation}
\overline{G(z_1)G(z_2)} =
\sum_{r=0}^{\infty} 
\sum_{s=0}^{\infty} 
{ 1 \over z_1^{r+1} z_2^{s+1} } \; \overline{\tr(H^r)\tr(H^s)} \ .
\label{r1}
\end{equation}
The ensemble average vanishes unless $r+s$ is even, and, in this case,
Wick's theorem can be used to express it as a sum over $(r+s-1)!!$ pairwise
contractions of $V$'s.  Each contraction that occurs
inside one of the two traces can be shown \cite{brw2} to yield
a factor of $v_{0k}^2 \Lambda_k$, where $\Lambda_k$ is a calculable number
(given below), 
provided the total number of these contractions is much less than 
$\ell$.  This is essence of the binary correlation method,
introduced by Mon and French \cite{mf} (see also \cite{vz}).
It is convenient to choose $v_{0k}$ so that
$v_{0k}^2 \Lambda_k = 1$; then we have
\begin{equation}
v_{0k}^{-2} = \Lambda_k = {m \choose k}{\ell-m+k \choose k} \ .
\label{v0k}
\end{equation}
It remains to count the total number
of contractions within each trace, and to evaluate the contractions
across the two traces.  Performing the first task yields \cite{brw2}
\begin{equation}
R(z_1,z_2)=
\sum_{n=1}^{\infty} g_n(z_1) g_n(z_2) T_n \ , 
\label{r2}
\end{equation}
where
\begin{equation}
g_n(z)=
\sum_{p=0}^{\infty} 
{ (2p-1)!! \over z^{2p+n+1} } {n+2p \choose n} 
\label{gn}
\end{equation}
and
\begin{equation}
T_n = \overline{\overline{\tr(H^n)\tr(H^n)}} \ ;
\label{tn}
\end{equation}
the double overbar means that all contractions of $V$'s are
to involve one $V$ from each of the two traces.
Eqs.~(\ref{r2}--\ref{tn}) are equivalent to eq.~(68)
of \cite{brw2}.

BRW show that $T_n\sim \ell^{-kn}$ for $\ell \gg m \gg k$, 
and so conclude that
$R(z_1,z_2)$ vanishes in the $\ell\to\infty$ limit;
this is indicative of Poissonian spectral statistics.
However, this conclusion is suspect if we do not first
understand the convergence properties
of the series in eq.~(\ref{r2}).
To see why this is necessary, consider the mathematical example
\begin{equation}
S = \sum_{n=1}^\infty {n!\over\ell^n} \ ,
\label{s}
\end{equation}
where $\ell$ is a positive integer.
Clearly, in the limit $\ell\to\infty$, each term of this series
vanishes, and so it is tempting to conclude that $S=0$.
However, we can evaluate $S$ by Borel summation.
We use
\begin{equation}
n! = \int_0^\infty dt\,e^{-t}\,t^n 
\label{n!}
\end{equation}
in eq.~(\ref{s}) and do the sum to get
\begin{equation}
S = \int_0^\infty dt\,e^{-t}\,{t\over{\ell-t}}\ .
\label{s2}
\end{equation}
This integral is not defined for positive real $\ell$.
We could attempt to define it by analytic continuation,
but there would still be an ambiguity, corresponding to
whether the positive real axis is approached from above
or below.
More importantly, in our case $\ell$ is a positive integer
that counts the number of single particle states;
therefore it does not seem to make sense to analytically
continue to complex $\ell$\footnote{One could raise
the same objection to dimensional regularization in 
quantum field theory, but there the results can be
verified by a variety of different and more physically
motivated schemes.}.  
Our conclusion in this case
would be that $S$ is simply not defined by the Borel
procedure.  We wish to examine whether the same problem
arises for the series in eq.~(\ref{r2}).

The series for $g_n(z)$ in eq.~(\ref{gn}), on the other hand,
can be defined for any $z$ with $\im z \ne 0$
by Borel summation followed by 
analytic continuation.  To demonstrate this,
we begin with the combinatoric identities
\begin{equation}
{n+2p \choose n} = {(n+2p)!\over n!(2p)!} \qquad {\rm and}
\qquad (2p-1)!!={(2p)!\over 2^p p!} \ ,
\label{comb}
\end{equation}
and so 
\begin{equation}
g_n(z)={1\over n!\,z^{n+1}}
\sum_{p=0}^{\infty} 
\left({1\over 2z^2}\right)^{\!\! p} {(n+2p)!\over p!} \ .
\label{g2}
\end{equation}
We now use eq.~(\ref{n!}) with $n\to n+2p$ to get
\begin{eqnarray}
g_n(z) &=& {1\over n!\,z^{n+1}}
\sum_{p=0}^{\infty} 
\int_0^\infty dt\,e^{-t}
\left({t^2\over 2z^2}\right)^{\!\! p} {t^n\over p!} 
\nonumber \\
&=& {1\over n!\,z^{n+1}}\int_0^\infty dt\,e^{-t}\,t^n\,e^{t^2\!/2z^2} \ .
\label{g22} 
\end{eqnarray}
The integral converges provided $\re z^{-2} < 0$.
This condition is satisfied for $z=\pm i|z|e^{i\phi}$ with
$-{1\over4}\pi<\phi<+{1\over4}\pi$.  We now rotate the $t$ contour
in the complex $t$ plane so that it runs along a straight
line from zero to complex infinity at an angle of $\phi$ 
relative to the positive real axis.  Then we set $t=\tau e^{i\phi}$,
where $\tau$ is real and runs from zero to infinity.  We now have
\begin{equation}
g_n(z)={e^{i(n+1)\phi}\over n!(\pm i)^{n+1}|z|^{n+1}e^{i(n+1)\phi}}
\int_0^\infty d\tau\,e^{-\tau e^{i\phi}}\tau^n e^{-\tau^2/2|z|^2} \ ,
\label{g3} 
\end{equation}
where the phase in the numerator of the prefactor comes from
the change of variable $t\to\tau e^{i\phi}$, and the 
phase in the denominator comes from $z=\pm i|z|e^{i\phi}$.
We now change the integration variable to $u=\tau/|z|$ to get 
\begin{eqnarray}
g_n(z) &=& {(\mp i)^{n+1}\over n!}
\int_0^\infty du\,e^{-u|z|e^{i\phi}}u^n\,e^{-u^2/2} 
\nonumber \\
&=& {(\mp i)^{n+1}\over n!}
\int_0^\infty du\,e^{\pm i z u}\,u^n\,e^{-u^2/2} \ . 
\label{g4} 
\end{eqnarray}
This integral converges for all $z$, and so constitutes
an analytic continuation of eq.~(\ref{g22});
the $\pm$ symbol should be interpreted as the sign
of $\im z$.  Thus $g_n(z)$ is discontinuous
across the real axis, and if we take
$z = z^\pm = E \pm i\varepsilon$, we get
\begin{equation}
g_n(z^\pm) = {(\mp i)^{n+1}\over n!}
\int_0^\infty du\,e^{\pm i E u}\,u^n\,e^{-u^2/2} \ . 
\label{g5} 
\end{equation}
We also note that, from our eq.~(\ref{gn})
and eq.~(65) of \cite{brw2}, 
\begin{equation}
\overline{G(z)} = g_0(z) \ .
\label{g6} 
\end{equation}
Then, from eq.~(\ref{rho}) and eq.~(\ref{g5}), we have
\begin{eqnarray}
\overline{\rho(E)} &=& {1\over2\pi i}\Bigl[g_0(z^-)-g_0(z^+)\Bigr]
\nonumber \\
&=& 
{1\over 2\pi}
\int_0^\infty du\,(e^{-iEu}+e^{+iEu})\,e^{-u^2/2}  
\nonumber \\
&=& 
{1\over\sqrt{2\pi}}\exp(-E^2\!/2) \ .
\label{rho2}
\end{eqnarray}
This is the classic result of Mon and French \cite{mf}: 
the ensemble-averaged density of states of the 
$k$-body random-interaction model is Gaussian.

Returning to $R(z_1,z_2)$, we use eqs.~(\ref{r2}) and
(\ref{g4}) to get
\begin{equation}
R(z_1^+,z_2^\pm)= \mp \int_0^\infty du\;dv\;
                      e^{i(E_1 u\pm E_2 v)}\,
                      e^{-(u^2+v^2)/2}
                      F(\mp u v) \ ,
\label{r3}
\end{equation}
where we have defined
\begin{equation}
F(y) \equiv \sum_{n=1}^\infty {y^n\over(n!)^2}\,T_n \ .
\label{f}
\end{equation}
We see that,
in order for both $R(z_1^+,z_2^+)$ and $R(z_1^+,z_2^-)$
to be well defined, $F(y)$ must be free of singularities
on the real axis (positive or negative).  Only then
can we perform the integrals over $u$ and $v$ in eq.~(\ref{r3})
for both $R(z_1^+,z_2^+)$ and $R(z_1^+,z_2^-)$
without further (arbitrary) regularization.

To see whether or not this obstacle arises, we must evaluate $T_n$.
Here we have an immediate difficulty.  We have already
invoked the binary correlation approximation in eq.~(\ref{gn});
the terms in this series receive corrections when the summation
index $p$ becomes comparable to $\ell$, and there is no 
straightforward way to calculate these corrections exactly.
Similarly, in section~3 we will evaluate $T_n$ (for $k=1$), 
but our method will require $n\ll\ell$.  It is therefore
unsuitable for reliably determining the asymptotic behavior
of the series in eq.~(\ref{f}).  Still, it is worthwhile
to see whether or not the problem of a singularity on the
real axis arises within this approximation.
We therefore turn to the calculation of $T_n$ for $k=1$.

\section{Analysis for $k=1$}

We specialize to the case $k=1$ (and drop the corresponding
``1'' subscripts):
\begin{equation}
H = \sum_{j,i} V_{ji}\,a_j^{\dagger}a_i \ ,
\label{h1}
\end{equation}
with
\begin{equation}
\overline{V_{ji} V_{j'i'}}=v_0^2\,\delta_{ji'}\delta_{ij'} \, .
\label{v1}
\end{equation}
Of course, we already know the answer for this case:
the spectral statistics are Poissonian, because the spectrum
simply consists of the linear sum of the $m$ single-particle energies
that are obtained by diagonalizing the $\ell\times\ell$ hermitian
matrix $V$.
These single-particle energies obey Wigner-Dyson statistics, but
their sums (for $m\gg 1$) obey Poisson statistics.  
However, the binary correlation method can still be used,
and it is important to see whether or not it gives the correct
answer (or any answer at all).  

We first introduce some shorthand notation.  
Let $I=\{i_1,\ldots,i_n\}$ and
$J=\{j_1,\ldots,j_n\}$; let
\begin{equation}
\delta_{I,J} = \delta_{i_1 j_1} \ldots \delta_{i_n j_n} \ .
\label{delta}
\end{equation}
Let $P$, $Q$, and $R$ denote permutations
of the $n$ indices of $I$.  Then, from eq.~(\ref{v0}),
we can write
\begin{equation}
\overline{\overline{
(V_{j_1 i_1} \ldots V_{j_n i_n})
(V_{j'_1 i'_1} \ldots V_{j'_n i'_n}) }}
=v_0^{2n} \sum_R \delta_{I,RJ'} \delta_{J,RI'}
\label{v2}
\end{equation}
Another useful bit of shorthand is
\begin{equation}
A^\dagger_J A_I = a^\dagger_{j_1}a_{i_1}\ldots a^\dagger_{j_n}a_{i_n}\ .
\label{AdA}
\end{equation}
Thus eq.~(\ref{tn}) becomes
\begin{equation}
T_n = 
v_0^{2n} \sum_R \sum_{IJI'J'} \delta_{I,RJ'} \delta_{J,RI'}
\tr(A^\dagger_J A_I)
\tr(A^\dagger_{J'} A_{I'}) \ .
\label{tn2}
\end{equation}
The trace is a $U(\ell)$ invariant operation, and so we must have
\begin{equation}
\tr(A^\dagger_J A_I) = \sum_P C_P\,\delta_{J,PI} 
\label{CP}
\end{equation}
for some set of coefficients $C_P$.
Substituting this expansion into eq.~(\ref{tn2}), we get
\begin{eqnarray}
T_n &=& 
v_0^{2n} \sum_{P,Q,R}C_P C_Q \sum_{IJI'J'} \delta_{I,RJ'} \delta_{J,RI'}
                                          \delta_{J,PI} \delta_{J',QI'} 
\nonumber \\
&=& v_0^{2n} \sum_{P,Q,R}C_P C_Q \sum_I \delta_{I,RQ\overline{R}PI} \,
\label{tn3}
\end{eqnarray}
where $\overline{R}$ is the inverse of $R$.  

The expression $\sum_I \delta_{I,PI}$
can be evaluated in terms of the cycles of $P$.
Consider, for example, the permutation
2431 of 1234; it contains one cycle of length one
(since 3 remains in its original position),
and one cycle of length three
(since 1 is replaced by 2, 2 is replaced by 4,
and 4 is replaced by 1).  Let $\gamma_c(P)$ denote
the number of cycles of length $c$ in permutation $P$;
for our example, $\gamma_1=\gamma_3=1$ 
and $\gamma_2=\gamma_4=0$.
Since each element of any permutation
is in exactly one cycle, we have
$\sum_{c=1}^n \gamma_c(P)c = n$.
In $\sum_I \delta_{I,PI}$, each cycle of $P$ ultimately
results in a factor of $\sum_i \delta_{ii}=\ell$.
Thus we have
\begin{equation}
\sum_I \delta_{I,PI} = \ell^{\,\sum_c\!\gamma_c(P)} \ ,
\label{dIPI}
\end{equation}
and hence
\begin{equation}
T_n = v_0^{2n} \sum_{P,Q,R}C_P C_Q \,
               \ell^{\,\sum_c\!\gamma_c(RQ\overline{R}P)} \ .
\label{tn4}
\end{equation}

So far we have made no approximations in our evaluation of $T_n$.
We now notice that, for $\ell \gg 1$, the dominant term on the 
right-hand side of eq.~(\ref{tn4}) is the one with the largest value of 
$\sum_c \gamma_c(RQ\overline{R}P)$.  This occurs when 
$\gamma_1=n$ and $\gamma_c=0$ for $2\le c\le n$, 
which in turn implies $RQ\overline{R}P=I$, or 
$Q=\overline{R}\,\overline{P}R$.  
Thus we have
\begin{equation}
T_n \cong v_0^{2n}\,\ell^n 
          \sum_{P,R} C_P C_{\overline{R}\,\overline{P}R} \ ,
\label{tn6}
\end{equation}
where $\cong$ denotes equality in the limit of large $\ell$.

Next we must evaluate $C_P$.
Starting with eq.~(\ref{CP}), we set $J=QI$ and sum over $I$ 
to get
\begin{eqnarray}
\sum_I \tr(A^\dagger_{QI} A_I) 
&=& \sum_P C_P \sum_I\delta_{QI,PI}
\nonumber \\
&=& \sum_P C_P \,\ell^{\,\sum_c\!\gamma_c(\overline{Q}P)} 
\nonumber \\
&\cong& C_Q\,\ell^n \ . 
\label{CP2}
\end{eqnarray}
To get the last line, we used the same large-$\ell$ argument
that gave us eq.~(\ref{tn6}).

Now we need to evaluate 
$\sum_I\tr(A^\dagger_{PI} A_I)$.
To do so, imagine
rearranging the $a^\dagger$'s so that their indices
are in the same order as the $a$'s, and interleaved among
them in the standard pattern of eq.~(\ref{AdA}).
Since
$a^\dagger_i a^\dagger_j = - a^\dagger_j a^\dagger_i$,
we see that, before accounting for the presence of the $a$'s,
rearranging the $a^\dagger$'s in this way will yield
a factor of $(-1)^P$.
Passing $a^\dagger$'s through $a$'s yields Kronecker deltas,
since $a_i a^\dagger_j = -(a^\dagger_j a_i - \delta_{ij})$,
but no possibility of an extra overall minus sign,
since the total number of exchanges of $a^\dagger$'s with $a$'s
is always even.  Let us temporarily ignore the Kronecker deltas.  
Then, once all the $a^\dagger$'s have been put 
next to their partner $a$'s in the standard pattern,
we have $\sum_i a^\dagger_i a_i=m$.
Thus, ignoring the Kronecker deltas generated by moving $a^\dagger$'s
past $a$'s, we have $C_P = (-1)^P m^n$.

For $n\ll\ell$,
most of the Kronecker delta's can indeed be neglected; the exception
occurs when one of them has two identical indices, leading to 
$\sum_i\delta_{ii}=\ell$.
To account for these dominant Kronecker deltas, we first note that,
for a given permutation $P$, $\gamma_1(P)$ is the number
of elements that are left in the same place by $P$,
and these contribute no Kronecker deltas.
Let $r(P)$ be the number of elements that are moved to the right of their
original positions by $P$; only these $a^\dagger$'s 
need to pass through their partner $a$'s on being
returned to the standard order.  For these, the factor of $m$ must
be replaced by $m-\ell$.  Finally, the number of elements that are 
moved to the left of their original positions by $P$ is
$n-\gamma_1(P)-r(P)$, and these contribute no large Kronecker deltas
on being returned to the standard order.  Thus we have
\begin{equation}
C_P \cong (-1)^P\,\ell^{-n}\, m^{\gamma_1(P)}
                           \, m^{n-\gamma_1(P)-r(P)}
                           \, (m-\ell)^{r(P)} 
\label{CP4}
\end{equation}
for $n \ll \ell$.

Next we notice
that the inverse permutation $\overline P$ has the counts of the left- and 
right-moving elements exchanged.  Thus,
\begin{equation}
C_{\overline P} \cong (-1)^P\,\ell^{-n}\,m^{\gamma_1(P)}
                                       \,m^{r(P)} 
                                       \,(m-\ell)^{n-\gamma_1(P)-r(P)} \ ,
\label{CP5}
\end{equation}
Then, using the fact that $(-1)^P$ and $\gamma_1(P)$ are each
 invariant under unitary transformations of $P$, we have
\begin{equation}
C_{\overline{R}\,\overline{P}R} \cong (-1)^P\, \ell^{-n}\,
                      m^{\gamma_1(P)} \,
                      m^{r(\overline{R}PR)} \,
                      (m-\ell)^{n-\gamma_1(P)-r(\overline{R}PR)} \ .
\label{CP6}
\end{equation}
Putting eqs.~(\ref{CP4}) and (\ref{CP6}) into eq.~(\ref{tn6}), we get
\begin{equation}
T_n \cong v_0^{2n} \,[m(m-\ell)]^n \,\ell^{-n}
                 \sum_{P,R} \left({m\over m-\ell}
                 \right)^{\gamma_1(P)-r(P)+r(\overline{R}PR)} \ .
\label{tn7}
\end{equation}
The convention of eq.~(\ref{v0k}) with $k=1$ yields
$v_0^{2n}[m(m-\ell)]^n = (-1)^n$.  Then we can write
\begin{equation}
T_n/(n!)^2 \cong \ell^{-n} M_n(f/(1{-}f)) \ ,
\label{tn8}
\end{equation}
where $f=m/\ell$ is the filling fraction,
and we have defined the function
\begin{equation}
M_n(x) \equiv {(-1)^n\over(n!)^2}\sum_{P,R}(-x)^{\gamma_1(P) - r(P)
                                    + r(\overline{R}PR)} \ .
\label{M}
\end{equation}
Remarkably, it is possible to evaluate $M_n(x)$ exactly for
arbitrary $n$ without performing the sum over the permutations
$P$ and $R$ in eq.~(\ref{M}) explicitly.  This calculation is given in
the Appendix.  We find that $M_n(x)$ is everywhere positive
and convex; it diverges as $x\to0$ and $x\to+\infty$, and has 
a minimum between $x=0$ and $x=1$.  While we are not able to evaluate the
asymptotic form of $M_n(x)$ analytically, numerical evaluation
reveals that, for large $n$, 
\begin{equation}
M_n(x) \sim n\lambda^n \ , 
\label{M2}
\end{equation}
where $\lambda$ is a number that depends on $x$;
if $x$ is real and positive, so is $\lambda$.
Using eqs.~(\ref{tn8}) and (\ref{M2}) in eq.~(\ref{f}), we find
\begin{equation}
F(y) \sim { \lambda y/\ell \over (1 - \lambda y/\ell)^2 } \ ,
\label{f2}
\end{equation}
which obviously has a singularity on the positive real axis
at $y=+\ell/\lambda$.
Hence the integral in eq.~(\ref{r3}) for $R(z_1^+,z_2^-)$ 
is not well defined.  Therefore, we argue, the binary correlation method
fails to give us the density-density correlation function.

\section{Conclusions}

It is significant that, while we cannot compute $R(z_1^+,z_2^-)$
by the binary correlation method, we {\it can\/} compute $R(z_1^+,z_2^+)$.
In this case (at least for $k=1$), 
there is no singularity in the integrand in eq.~(\ref{r3}),
and so we can complete the integrals, take the $\ell\to\infty$ limit,
and conclude that $R(z_1^+,z_2^+)=R(z_1^-,z_2^-)^*=0$.
This is consistent with other approaches to the computation of
this spectral correlation function.  For example, in the periodic-orbit
approach to chaotic systems, one finds (see, e.g., \cite{bogo})
\begin{equation}
G(z^\pm) = \sum_{\rm p.o.} w_p \,e^{\mp i S_p/\hbar}e^{-\varepsilon T_p} \ ,
\label{gutz}
\end{equation}
where $S_p$ is the action of the orbit (including the contribution
of the Maslov phase), $T_p$ is its period, and 
$w_p$ is the Gutzwiller weight factor.
The key point is that, in the limit of small $\hbar$,
$G(z_1^+)G(z_2^+)$ contains only large ($\gg 2\pi$) phases, 
and so summing the orbits should give zero.
On the other hand, $G(z_1^+)G(z_2^-)$
has negative phases from the first factor and positive from the second,
so cancellations can occur; in the most straightforward approach
(the so-called diagonal approximation), one simply keeps only those
terms with an exact cancellation.  This is adequate for computing
the density-density correlation function at large $|E_1-E_2|$;
for chaotic systems, it is nonzero, and agrees with the Wigner-Dyson
prediction of random-matrix theory.  Thus, for the $k$-body 
random-interaction model, the binary-correlation
method correctly and unambiguously gives us $R(z_1^+,z_2^+)=0$,
but, as we have argued, formal manipulations of divergent series
are not adequate for extending this result to $R(z_1^+,z_2^-)$.
We have seen explicitly how the binary-correlation approximation
fails to give us a series for $R(z_1^+,z_2^-)$ that can be
evaluated by Borel summation in the case $k=1$.

It is important to remember, though, that the binary correlation
approximation breaks down precisely where we wish to apply it:
when the number of contractions approaches the number of single-particle
levels.  Thus, to reach a firm conclusion one way or the other, we
are forced to go beyond this approximation.  At the present time, 
doing so appears to present a severe challenge to the available
technologies.

One of these technologies, also investigated by BRW, is the supersymmetric
sigma model \cite{ef}.  BRW found that, at the tree and one-loop
levels, the sigma model predicts Wigner-Dyson statistics for
$k\ll m \ll \ell$, with corrections that go to zero as $1/\ell$.  
Unfortunately, this does not settle the matter, because
this prediction also applies to $k=1$, where it is known to be false.
One must therefore presume that higher-loop corrections are important.

Thus the only conclusion that seems completely safe at this juncture
is a disappointing one: the nature of the spectral statistics of the
$k$-body random-interaction model remains an unsolved problem.

\vskip0.2in

I am grateful to Hans Weidenm\"uller and (especially) Luis Benet
for open and extensive discussions of their work.  I also thank
Felix Izrailev, Francois Leyvraz, and Thomas Seligman for discussions, 
and Steve Tomsovic for help with the combinatorics.
I also thank Thomas Seligman for the kind hosptialisty of the 
Centro Internacional de Ciencias A.C., where this work was begun 
under the additional sponsorship of CONACyT--Mexico.
This work was also supported in part by NSF Grant PHY99-70701.

\section{Appendix}

We wish to evaluate
\begin{equation}
(-1)^n (n!)^2 M_n(-x) = 
\sum_{P,R}x^{\gamma_1(P) - r(P) + r(\overline{R}PR)} \ ,
\label{M3}
\end{equation}
where $P$ and $R$ are permutations of $n$ elements,
$\gamma_1(P)$ is the number of elements left in their
original positions by $P$, and $r(P)$ is the number of elements
that are moved to the right of their original positions by $P$.
For small values of $n$, we can perform this sum over $(n!)^2$
terms explicitly.  We are, however, interested in its behavior
for large $n$, and so this strategy quickly becomes untenable.

Let us warm up by computing 
\begin{equation}
R_n(x) = \sum_P x^{r(P)} \ .
\label{rn}
\end{equation}
Let $N_{n,r}$ be the number of permutations of $n$ elements
in which exactly $r$ of the elements move to the right.  Then
\begin{equation}
R_n(x) = \sum_{r=0}^{n-1}N_{n,r} x^r \ .
\label{rn2}
\end{equation}
We can compute $N_{n,r}$ by the method of inclusion and exclusion \cite{r}.
Let $S_{n,r}$ be the number of permutations in which {\em at least\/}
$r$ elements move to the right.  Then \cite{r}
\begin{equation}
N_{n,r} = \sum_{k=0}^{n-r} (-1)^k {r+k \choose k} S_{n,r+k} \ .
\label{nnr}
\end{equation}
To obtain $S_{n,r}$, let us first count the number of permutations
in which $r$ {\em specific\/} elements move to the right, irrespective
of what happens to the remaining elements.  Actually,
since we conventionally count from left to right, it is slightly more
calculationally convenient to count the number of permutations that move
$r$ specific elements to the left, rather than to the right; clearly, by
left-right symmetry, the choice of direction is irrelevant to the count.
So, let $A_{n,i_1}$ be the
number of permutations in which the $i_1^{\rm th}$ element moves to the left.
There are $i_1-1$ possible places for the $i_1^{\rm th}$ element to go, 
and $(n-1)!$ permutations of the remaining elements; thus
\begin{equation}
A_{n,i_1} = (i_1-1)(n-1)! \ .
\label{an}
\end{equation}
Similarly, let $A_{n,i_1 \ldots i_r}$, with $i_1<\ldots<i_r$, be the number
of permutations in which all the named elements move to the left; by a
similar argument, we have
\begin{equation}
A_{n,i_1 \ldots i_r} = (i_1-1)(i_2-2)\ldots(i_r-r)(n-r)! \ .
\label{an2}
\end{equation}
Now $S_{n,r}$ is given by the sum over all possible values of the $i$'s of 
$A_{n,i_1 \ldots i_r}$:
\begin{eqnarray}
S_{n,r} &=& \sum_{i_r=r}^n \ldots 
            \sum_{i_2=2}^{i_3-1}
            \sum_{i_1=1}^{i_2-1} 
            A_{n,i_1 \ldots i_r}
\nonumber \\
&=& (n-r)! \sum_{i_r=r}^n (i_r-r) \ldots 
           \sum_{i_2=2}^{i_3-1} (i_2-2)
           \sum_{i_1=1}^{i_2-1} (i_1-1)
\nonumber \\
&=& (n-r)! \sum_{j_r=0}^{n-r} j_r \ldots 
           \sum_{j_2=0}^{j_3} j_2
           \sum_{j_1=0}^{j_2} j_1 \ ,
\label{snr}
\end{eqnarray}
where, in the last line, $j_a = i_a-a$.
Eqs.~(\ref{rn2}), (\ref{nnr}), and (\ref{snr}) give us $R_n(x)$.
The first few of these polynomials are
\begin{eqnarray}
R_1(x) &=& 1
\nonumber \\
R_2(x) &=& 1 + x 
\nonumber \\
R_3(x) &=& 1 + 4x + x^2 
\nonumber \\
R_4(x) &=& 1 + 11x + 11x^2 + x^3 
\nonumber \\
R_5(x) &=& 1 + 26x + 66x^2 + 26x^3 + x^4 \ . 
\label{rs}
\end{eqnarray}
Also, it turns out to be convenient later if we adopt the convention
\begin{equation}
R_0(x) = 1/x \ .
\label{r0}
\end{equation}

Now consider $\sum_R x^{r(\overline R P R)}$ for fixed $P$.
Let $P$ belong to cycle class $\cal C$, specified by the numbers
$\gamma_c$ of cycles of length $c$, with $\sum_{c=1}^n c\gamma_c = n$.
The permutation $\overline R P R$ belongs to the same class
(since class is preserved by unitary transformations), and furthermore
each permutation in the class appears equally often as $R$ is varied
over the group.  Thus
\begin{equation}
\sum_R x^{r(\overline R P R)} = {n!\over N_{\cal C}}
\sum_{P\in{\cal C}} x^{r(P)} \ ,
\label{sumr}
\end{equation}
where
\begin{equation}
N_{\cal C} = {n! \over \prod_{c=1}^n c^{\gamma_c}\gamma_c!}
\label{nc}
\end{equation}
is the number of permutations in class $\cal C$ \cite{r}.
Let ${\widetilde N}_{c,r}$ be the number of permutations of
the elements in a particular cycle of length $c$ in which
exactly $r$ of the elements in this cycle move to the right.
A cycle can be specified by an ordering of its elements 
with the smallest first;
thus a cycle of length $c$ can be mapped to a permutation
of the remaining $c-1$ elements.  This mapping can be shown
to imply ${\widetilde N}_{c,r} = N_{c-1,r-1}$.  Thus,
\begin{eqnarray}
\sum_{r=1}^{c-1} {\widetilde N}_{c,r}x^r
&=& \sum_{r=1}^{c-1} N_{c-1,r-1}x^r
\nonumber \\
&=& \sum_{s=0}^{c-2} N_{c-1,s}x^{s+1}
\nonumber \\
&=& x R_{c-1}(x) \ .
\label{tilden}
\end{eqnarray}
The convention of eq.~(\ref{r0}) correctly treats the case $c=1$.
Also, the number of ways of assigning elements to the cycles is
\begin{equation}
N_A = {n! \over \prod_{c=1}^n (c!)^{\gamma_c}\gamma_c!} \ .
\label{na}
\end{equation}
Thus
\begin{eqnarray}
\sum_{P\in{\cal C}} x^{r(P)} &=& N_A \prod_{\rm cycles} x R_{c-1}(x)
\nonumber \\
&=& N_A \prod_{c=1}^n \Bigl[xR_{c-1}(x)\Bigr]^{\gamma_c}
\nonumber \\
&=& n! \prod_{c=1}^n{1\over\gamma_c!}
       \left[{xR_{c-1}(x)\over c!}\right]^{\gamma_c} \ .
\label{pinc}
\end{eqnarray}
Therefore, by eqs.~(\ref{sumr}), (\ref{nc}), and (\ref{pinc}),
\begin{equation}
\sum_R x^{r(\overline R P R)} = n! \prod_{c=1}^n
       \left[{xR_{c-1}(x)\over (c-1)!}\right]^{\gamma_c} \ .
\label{sumr2}
\end{equation}

Now we can write
\begin{eqnarray}
(-1)^n (n!)^2 M_n(-x) &=&
\sum_{P,R}x^{\gamma_1(P) - r(P) + r(\overline{R}PR)} 
\nonumber \\
&=& \sum_{\cal C} \sum_{P\in\cal C}x^{\gamma_1(P)-r(P)}
    \sum_R x^{r(\overline R P R)} \ .
\label{M4}
\end{eqnarray}
The sum over $R$ is given by eq.~(\ref{sumr2}),
and the sum over 
$P\in{\cal C}$ is given by eq.~(\ref{pinc})
with $xR_{c-1}(x)$ replaced by
$x^{\delta_{c1}-1}R_{c-1}(1/x)$.
The sum over classes is equivalent to a sum over all possible
values of each $\gamma_c$ with the constraint that
$\sum_{c=1}^n c\gamma_c = n$.  This is most easily treated
via a generating function
\begin{equation}
{\cal M}(x,z) = \sum_{n=0}^\infty M_n(x)z^n \ .
\label{mgen}
\end{equation}
Then, from eqs.~(\ref{pinc}--\ref{mgen}) we have
\begin{eqnarray}
{\cal M}(x,z) &=& \sum_{\gamma_c=0}^\infty \prod_{c=1}^\infty
                  {1\over\gamma_c!} \left[
        {(-z)^c(-x)^{\delta_{c1}} R_{c-1}(-x) R_{c-1}(-1/x) \over 
                          c!(c-1)!}\right]^{\gamma_c} 
\nonumber \\
              &=& \exp\Biggl[\,\sum_{c=1}^\infty
                  {(-z)^c (-x)^{\delta_{c1}} 
                   R_{c-1}(-x) R_{c-1}(-1/x) \over 
                   c!(c-1)!}\Biggr]  \ .
\label{mgen2}
\end{eqnarray}
The expansion of eq.~(\ref{mgen2}) in powers of $z$ yields
$M_n(x)$ as the coefficient of $z^n$.  
The results for small $n$ are
\begin{eqnarray}
M_1(x) &=& x
\nonumber \\
M_2(x) &=& (1 + x^2)/2
\nonumber \\
M_3(x) &=& ( 1/x - 2 + 7x + 2x^3 )/(2!3!)
\nonumber \\
M_4(x) &=& 
( 1/x^2 - 8/x + 48 - 32x + 49x^2 + 6x^4 )/(3!4!) \ .
\label{ms}
\end{eqnarray}
We have verified this procedure for computing $M_n(x)$ 
up through $n=7$ by comparison with the brute-force
summation of the right-hand side of eq.~(\ref{M3}).

It is tempting to conclude from eq.~(\ref{ms}) that,
in the limit of small $x$, we can take 
$M_n(x) = x^{-(n-2)}/(n!(n{-}1)!)$.  Small $x$ corresponds
to small filling fraction $f=m/\ell$, which is what we want
if we are interested in the limit $\ell\to\infty$ with
$m$ fixed.  This limiting form for $M_n(x)$ would lead
to convergence of the series in eq.~(\ref{f})
for all $y$, with no singularities on the real axis,
thus apparently validating the BRW result.
However, including the first correction yields
\begin{equation}
M_n(x) = x^{-(n-2)} 
         \Bigl[1 - (2^n{-}2n)x + \ldots\Bigr]/
         (n!(n{-}1)!) \ ,
\label{ma}
\end{equation}
and so we see that we must have $x \ll 2^{-n}$ 
before the first term alone is an adequate approximation.
Instead, for the purpose of determining the convergence
of the series in eq.~(\ref{f}), we should take 
the limit of large $n$ with $x$ held fixed.  
We have not found a way to do this
analytically, and so have resorted to numerical methods.
For a given value of $x$, we can evaluate $M_n(x)$ up to
around $n=500$ from eq.~(\ref{mgen2}) in a reasonable
amount of computation time.  We find, for $x=0.001$, 0.01,
0.1, 0.5, and 1.0, that the behavior of $M_n(x)$ for 
$n>10$ is very well fit by eq.~(\ref{M2}),
with $\lambda=17.4$, 3.28, 0.797, 0.435, and 0.405,
respectively.

\end{document}